\documentclass[USenglish,twocolumn]{article}

\usepackage[utf8]{inputenc}				
\usepackage[big,online]{dgruyter}	
\usepackage{lmodern}
\usepackage{microtype}
\usepackage[numbers,square,sort&compress]{natbib}
\usepackage{amsmath}
\usepackage{amsfonts}
\usepackage{graphicx}
\usepackage{epsfig}
\usepackage{color}

\theoremstyle{dgthm}

\theoremstyle{dgdef}

\begin{document}

  \articletype{Research Article}
  \received{Month	DD, YYYY}
  \revised{Month	DD, YYYY}
  \accepted{Month	DD, YYYY}
  \journalname{De~Gruyter~Journal}
  \journalyear{YYYY}
  \journalvolume{XX}
  \journalissue{X}
  \startpage{1}
  \aop
  \DOI{10.1515/sample-YYYY-XXXX}

\title{Vector optomechanical entanglement}
\runningtitle{Short title}

\author[2]{Ying Li}
\author[2]{Ya-Feng Jiao}
\author[3]{Jing-Xue Liu}
\author[4]{Adam Miranowicz}
\author[3]{Yun-Lan Zuo}
\author*[1]{Le-Man Kuang}
\author*[1]{Hui Jing}
\runningauthor{Y.~Li et al.:Vector optomechanical entanglement}
\affil[1]{\protect\raggedright
Key Laboratory of Low-Dimensional Quantum Structures and Quantum Control of Ministry of Education, Department of Physics and Synergetic Innovation Center for Quantum Effects and Applications, Hunan Normal University, Changsha 410081, China, e-mail: lmkuang@hunnu.edu.cn (L. Kuang), jinghui73@foxmail.com (H. Jing)}
\affil[2]{\protect\raggedright
Key Laboratory of Low-Dimensional Quantum Structures and Quantum Control of Ministry of Education, Department of Physics and Synergetic Innovation Center for Quantum Effects and Applications, Hunan Normal University, Changsha 410081, China. These authors contributed equally to this work.}
\affil[3]{\protect\raggedright
	Key Laboratory of Low-Dimensional Quantum Structures and Quantum Control of Ministry of Education, Department of Physics and Synergetic Innovation Center for Quantum Effects and Applications, Hunan Normal University, Changsha 410081, China.}
\affil[4]{\protect\raggedright
Institution, Department, City, Country of second author and third author, Institute of Spintronics and Quantum Information, Faculty of Physics, Adam Mickiewicz University, 61-614 Pozna\'{n}, Poland}

	
\abstract{The polarizations of optical fields, besides field intensities, provide more degrees of freedom to manipulate coherent light-matter interactions. Here we propose how to achieve a coherent switch of optomechanical entanglement in a polarized-light-driven cavity system. We show that by tuning the polarizations of the driving field, the effective optomechanical coupling can be well controlled and, as a result, quantum entanglement between the mechanical oscillator and the optical transverse electric (TE) mode can be coherently and reversibly switched to that between the same phonon mode and the optical transverse magnetic (TM) mode. This ability of switching optomechanical entanglement with such a vectorial device can be important for building a quantum network being capable of efficient quantum information interchanges between processing nodes and flying photons.}

\keywords{polarization; cavity optomechanics; quantum entanglement.}

\maketitle

\section{Introduction}

Vector beams, characterized by the ability to tailor light by polarization control, are important for both fundamental researches and practical applications in optics and photonics~\cite{chen2018Vectorial,rosales-guzman2018Review,Goldberg2021}. Manipulating the polarization of vector beams, for examples, provides efficient ways to realize optical trapping or imaging~\cite{doi:10.1021/acs.nanolett.7b00676,Kozawa:10,Chen:13}, material processing~\cite{ahn2018Optically,liu2020Robust}, optical data storage~\cite{xian2020Segmenteda}, sensing~\cite{Neugebauer2016}, and nonlinearity enhancement~\cite{Zhang2018}. Compared with conventional scalar light sources, vector beams provide more degrees of freedom to regulate coherent light-matter interactions, that is, manipulating the coupling intensity by tuning spatial polarization distributions of an optical field~\cite{purdy2013Observation,li2020Polarimetric,zhan2009Cylindrical,wang2020Vectorial}. Thus, vector beams have been used in a variety of powerful devices, such as optoelectrical~\cite{sederberg2020Vectorized,malossi2021Sympathetic} or optomechanical systems~\cite{he2016Optomechanical,xu2020Spontaneous}, metamaterial structures~\cite{kan2020Metasurface,buddhiraju2020Nonreciprocal}, and atomic gases~\cite{zhu2020Dielectricb}. In a recent experiment, through vectorial polarization control, coherent information transfer from photons to electrons was demonstrated~\cite{sederberg2020Vectorized}. In the quantum domain, mature techniques have been developed for creating entangled photons with vector beams~\cite{dambrosio2016Entangled,villar2020Entanglement}, and by using the destructive interference of the two excitation pathways of a polarized quantum-dot cavity, unconventional photon blockade effect was observed very recently~\cite{snijders2018Observation}. We also note that the single- and multi-photon resources with well-defined polarization properties, which are at the core of chiral quantum optics, have also been addressed via metasurfaces~\cite{stav2018Quantum,Wang2018}. In addition, a recent experiment shows that the hybrid light–mechanical states with a vectorial nature could also be achieved by using levitated optomechanical system~\cite{Ranfagni2021}. However, as far as we know, the possibility of generating and switching macroscopic entanglement between light and motion via polarization control has not yet been explored.

A peculiar property of quantum entanglement, that measuring one part of the entangled elements allows to determine the state of the other, makes it a key resource for quantum technologies, ranging from quantum information processing~\cite{RevModPhys.81.865,PhysRevLett.109.013603} to quantum sensing~\cite{Fleury2015,Guo2019}. Recently, entanglement-based secure quantum cryptography has also been achieved at a distance over 1,120\,kilometres~\cite{Yin2020}. So far, quantum entanglement has been observed in diverse systems involving photons, ions, atoms, and superconducting qubits~\cite{Volz2006,Cleland2004}. Quantum effects such as entanglement has also been studied at macroscopic scales and even in biological systems~\cite{adams2020Quantum}. In parallel, the rapidly emerging field of cavity optomechanics (COM), featuring coherent coupling of motion and light~\cite{aspelmeyer2014Cavity,verhagen2012Quantumcoherent}, has provided a vital platform for engineering macroscopic quantum objects~\cite{moore2021Searching,Qin2021,Lecocq2015}. Very recently, quantum correlations at room temperature were even demonstrated between light and a $40\,\textrm{kg}$ mirror, circumventing the standard quantum limit of measurement~\cite{Yu2020Quantum}. Quantum entanglement between propagating optical modes, between optical and mechanical modes, or between massive mechanical oscillators have all been realized in COM systems~\cite{Riedinger2016,vitali2007Optomechanical,Jiao2020,Chen2020,palomaki2013Entangling,
barzanjeh2019Stationary,Lepinay2021,OckeloenKorppi2018,Riedinger2018,Liao2014,Liao2016}. In view of these rapid advances, COM devices have becomes one of the promising candidates to operate as versatile quantum nodes processing or interchanging information with flying photons in a hybrid quantum network.

Here, based on a COM system, we propose how to achieve a coherent switch of quantum entanglement of photons and phonons through polarization control. We show that the intracavity field intensity and the associated COM entanglement can be coherently manipulated by adjusting the polarization of a driving laser. This provides an efficient way to manipulate the light-motion coupling, which is at the core of COM-based quantum technologies. Besides the specific example of COM entanglement switch, our work can also serve as the first step towards making vectorial COM devices with various structured lights, such as Bessel-Gauss beams, cylindrical beams or the Poincar\'{e} beams~\cite{poshakinskiy2021optomechanical,chen2014Generation,chen2018Vectorial}, where the optical polarization distribution is spatially inhomogeneous. Our work can also be extended to various COM systems realized with e.g., cold atoms, magnomechanical devices, and optoelectrical
circuits~\cite{nambu2020Observation,hu2021Noiseless}.

\section{Vectorial quantum dynamics}
	
\begin{figure}[t]
    \centering
    \includegraphics[width=8cm]{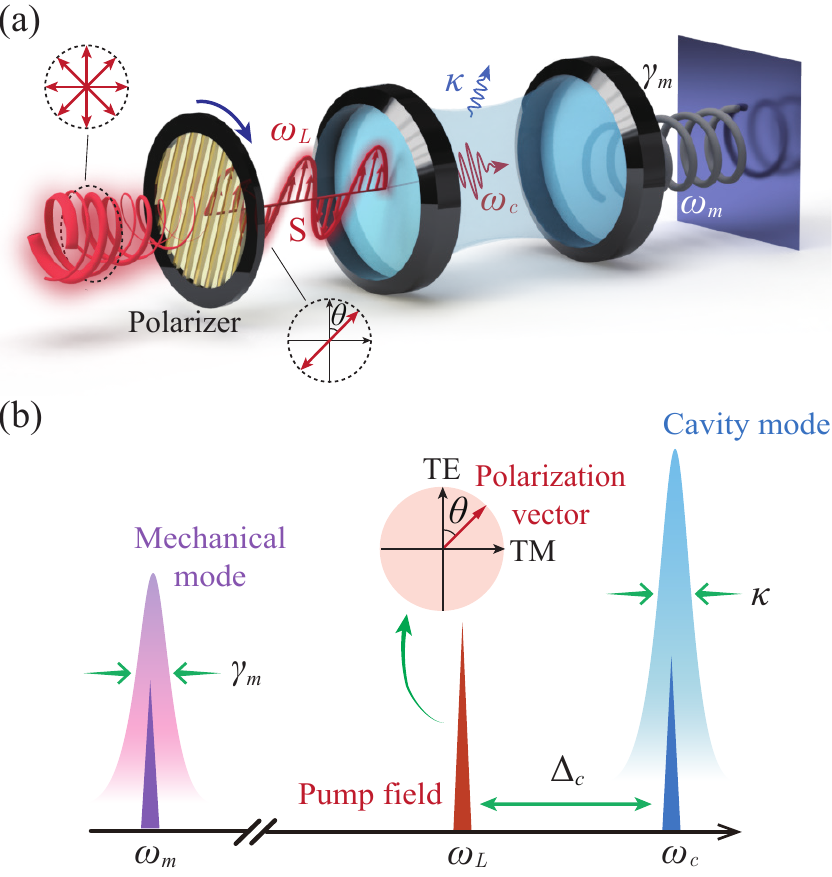}
     \caption{(a) Schematic diagram of a polarized-light-driven optomechanical system, which consists of an optical polarizer and a Fabry-P\'{e}rot cavity with a movable mirror. (b) Frequency spectrogram of a vector optomechanical system in panel (a), with $\theta$ being the angle between the polarization of the linearly polarized light and the vertical mode. The orthogonal cavity modes with degenerate resonance frequency $\omega_{c}$ and decay rate $\kappa$; the frequency of the mirror is $\omega_{m}$; the driving frequency is $\omega_{L}$, and $\Delta_{c}=\omega_{c}-\omega_{L}$ is optical detuning between the cavity mode and the driving field.}
	\label{FIG1}
\end{figure}

As shown in Figure\,\ref{FIG1}, we consider a polarized-light-driven optomechanical system, which consists of an optical polarizer and a Fabry-P\'{e}rot cavity with one movable mirror. Exploiting the polarization of photons rather than solely their intensity, has additional advantages for controlling light-matter interactions. Here, to describe the polarization of an optical field, it is convenient to introduce a set of orthogonal basis vectors, i.e., $|\vec{\textrm{e}}_{\updownarrow}\rangle$ and $|\vec{\textrm{e}}_{\leftrightarrow}\rangle$, which correspond to the vertical (TE) and horizontal (TM) modes of the Fabry-P\'{e}rot cavity~\cite{xiong2016Vector}. Therefore, an arbitrary linearly polarized light can be thought as a superposition of these orthogonal patterns, i.e., whose unit vector is represented by $|\vec{\textrm{e}}\rangle=\cos\theta\,|\vec{\textrm{e}}_{\updownarrow}\rangle+\sin\theta\,|\vec{\textrm{e}}_{\leftrightarrow}\rangle$, with $\theta$ being the angle between the polarization of the linearly polarized light and the vertical mode [see Figure\,\ref{FIG1}(b)]. In this situation, by adjusting the polarization angle $\theta$, one can coherently manipulate the spatial distribution of a linearly polarized optical field. In addition, because light could exert radiation pressure on the movable mirror, both spatial components of the linearly polarized light would experience an optomechanical interaction. Then, in a rotating frame with respect to $\hat{H}_{0}=\hbar\omega_{L}(\hat{a}_{\updownarrow}^{\dagger}\hat{a}_{\updownarrow}+\hat{a}_{\leftrightarrow}^{\dagger}\hat{a}_{\leftrightarrow})$, the Hamiltonian of the polarized-light-driven optomechanical system is given by
\begin{align}
&\hat{H}=\dfrac{\hbar\omega_{m}}{2}(\hat{p}^{2}\!+\!\hat{q}^{2})+\hbar\!\sum_{j=\updownarrow,\leftrightarrow}\left(\Delta_{c}\hat{a}_{j}^{\dagger}\hat{a}_{j}\!-\!g_{0}\hat{a}_{j}^{\dagger}\hat{a}_{j}\hat{q}\right)+\hat{H}_{\mathrm{dr}},
\nonumber\\
&\hat{H}_{\mathrm{dr}}=i\hbar\sqrt{2\kappa}\sum_{j=\updownarrow,\leftrightarrow}\left(\hat{a}_{j}^{\dagger}{S}_{j}-\hat{a}_{j}{S}_{j}^{\ast}\right),
\label{Ham}
\end{align}
where $\hat{a}_{j}$ $(\hat{a}_{j}^{\dagger})$ is the annihilation (creation) operator of the orthogonal cavity modes with degenerate resonance frequency $\omega_{c}$ and decay rate $\kappa$; $\hat{p}$ and $\hat{q}$ are, respectively, the dimensionless momentum and position operators of the mirror with mass $m$ and frequency $\omega_{m}$; $\Delta_{c}=\omega_{c}-\omega_{L}$ is optical detuning between the cavity mode and the driving field; $g_{0}$ is the single-photon optomechanical coupling coefficient for both orthogonal cavity modes. The frame rotating
with driving frequency $\omega_{L}$ is obtained by applying the unitary transformation $\hat{U}=\exp[i\hat{H}_{0}t/\hbar]$ (see, e.g., Ref.\,\cite{aspelmeyer2014Cavity}). $S=|S_{\updownarrow}|^2+|S_{\leftrightarrow}|^2=\sqrt{P/\hbar\omega_{L}}$ denotes the amplitude of a linearly polarized driving field with an input laser power $P$, where $S_{\updownarrow}=S\cos\theta$ and $S_{\leftrightarrow}=S\sin\theta$ are the projections of $S$ onto the vertical and horizontal modes, respectively.

By considering the damping and the corresponding noise term of both optical and mechanical modes, the quantum Langevin equations (QLEs) of motion describing the dynamics of this system are obtained as
\begin{align}
\dot{\hat{a}}_{j}&=(-i\Delta_{c}+ig_{0}\hat{q}-\kappa)\hat{a}_{j}
+\sqrt{2\kappa} S_{j}+\sqrt{2\kappa}\hat{a}_{j}^{\textrm{in}},\nonumber\\
\dot{\hat{q}}&=\omega_{m}\hat{p},\nonumber\\ \dot{\hat{p}}&=-\omega_{m}\hat{q}-\gamma_{m}\hat{p}+g_{0}\sum_{j=\updownarrow,\leftrightarrow}\hat{a}_{j}^{\dagger}\hat{a}_{j}+\hat{\xi},
\label{QLEs}
\end{align}
where $j=\updownarrow, \leftrightarrow$, $\gamma_{m}=\omega_{m}/Q_{m}$ is the mechanical damping rate with $Q_{m}$ the quality factor of the movable mirror; $\hat{a}_{j}^{\textrm{in}}$ are the zero-mean input vacuum noise operators for the orthogonal cavity modes, and they satisfy the following correlation functions~\cite{gardiner2004Quantum}
\begin{align}
&\langle\hat{a}_{j}^{\textrm{in},\dagger}(t)\hat{a}_{j}^{\textrm{in}}(t')\rangle
=0,\notag\\
&\langle\hat{a}_{j}^{\textrm{in}}(t)\hat{a}_{j}^{\textrm{in},\dagger}(t')\rangle
=\delta(t-t').
\label{opnof}
\end{align}
Moreover, $\hat{\xi}$ denotes the Brownian noise operator for the mechanical mode, resulting from the coupling of the mechanical mode with the corresponding thermal environment. It satisfies the following correlation function~\cite{giovannetti2001Phasenoise}
\begin{align}
\langle\hat{\xi}(t)\hat{\xi}(t')\rangle=\dfrac{\gamma_{m}}{\omega_{m}}\int\dfrac{d\omega}{2\pi}e^{-i\omega(t-t')}\left[\coth\left(\dfrac{\hbar\omega}{2k_{B}T}\right)+1\right],
\label{menof}
\end{align}
where $k_{B}$ is the Boltzmann constant and $T$ is the environment temperature of the mechanical mode. The noise operator $\hat{\xi}(t)$ models the mechanical Brownian motion as, in general, a non-Markovian process. However, in the limit of a high mechanical quality factor $Q_{m}\gg1$, $\hat{\xi}(t)$ can be faithfully considered as Markovian, and, then, its correlation function is reduced to
\begin{align}
&\langle\hat{\xi}(t)\hat{\xi}(t')\rangle\simeq\gamma_{m}(2n_{m}+1)\delta(t-t'),
\label{menoft}
\end{align}
where $n_{m}=[\textrm{exp}[(\hbar\omega_{m}/k_{B}T)]-1]^{-1}$ is the mean thermal phonon number.

Setting all the derivatives in QLEs~(\ref{QLEs}) as zero leads to the steady-state mean values of the optical and the mechanical modes
\begin{align}
\alpha_{j}=&\dfrac{\sqrt{2\kappa}}{i\Delta+\kappa}S_{j}~~
(j=\updownarrow, \leftrightarrow),\notag\\
q_{s}=&\dfrac{g_{0}}{\omega_{m}}\left(|\alpha_{\updownarrow}|^2+|\alpha_{\leftrightarrow}|^2\right), ~~
p_{s}=0,
\label{steadystate}
\end{align}
where $\Delta=\Delta_{c}-g_{0}q_{s}$ is the effective optical detuning.	Under the condition of intense optical driving, one can expand each operator as a sum of its steady-state mean value and a small quantum fluctuation around it, i.e., $\hat{a}_{j}=\alpha_{j}+\delta\hat{a}_{j}, \hat{q}=q_{s}+\delta\hat{q}, \hat{p}=p_{s}+\delta\hat{p}$. Then, by defining the following vectors of quadrature fluctuations and corresponding input noises : $u(t)=(\delta\hat{X}_{\updownarrow}, \delta\hat{Y}_{\updownarrow}, \delta\hat{X}_{\leftrightarrow}, \delta\hat{Y}_{\leftrightarrow}, \delta\hat{q}, \delta\hat{p})^{\textit{T}}$, $ v(t)=(\sqrt{2\kappa}\hat{X}_{\updownarrow}^{\textrm{in}}, \sqrt{2\kappa}\hat{Y}_{\updownarrow}^{\textrm{in}}, \sqrt{2\kappa}\hat{X}_{\leftrightarrow}^{\textrm{in}}, \sqrt{2\kappa}\hat{Y}_{\leftrightarrow}^{\textrm{in}}, 0,\hat{\xi})^{\textit{T}}$, with the components:
\begin{align}
\delta\hat{X}_{j}&=\dfrac{1}{\sqrt{2}}(\delta\hat{a}_{j}^{\dagger}
+\delta\hat{a}_{j}),& \delta\hat{Y}_{j}&=\dfrac{i}{\sqrt{2}}
(\delta\hat{a}_{j}^{\dagger}-\delta\hat{a}_{j}),
\nonumber\\
\hat{X}_{j}^{\textrm{in}}&=\dfrac{1}{\sqrt{2}}
(\hat{a}_{j}^{\textrm{in}\dagger}+\hat{a}_{j}^{\textrm{in}}),
&\hat{Y}_{j}^{\textrm{in}}&=\dfrac{i}{\sqrt{2}}(\hat{a}_{j}^{\textrm{in}\dagger}
-\hat{a}_{j}^{\textrm{in}}),
\label{quadraturefluctuation}
\end{align}
one can obtain a set of linearized QLEs, which can be written in a compact form as
\begin{align}
\dot{u}(t)=Au(t)+v(t),\label{lqles}
\end{align}
where
\begin{equation}
A=\begin{pmatrix}
-\kappa & \Delta & 0 & 0 & -G_{\updownarrow}^{y} & 0\\
-\Delta & -\kappa & 0 & 0 & G_{\updownarrow}^{x} & 0\\
0 & 0 & -\kappa & \Delta & -G_{\leftrightarrow}^{y} & 0\\
0 & 0 & -\Delta & -\kappa & G_{\leftrightarrow}^{x} & 0\\
0 & 0 & 0 & 0 & 0 & \omega_{m}\\
G_{\updownarrow}^{x} & G_{\updownarrow}^{y} & G_{\leftrightarrow}^{x} & G_{\leftrightarrow}^{y} & -\omega_{m} & -\gamma_{m}
\end{pmatrix}.
\label{amatrix}
\end{equation}
The linearized QLEs indicate that the effective COM coupling rate $G_{j}\equiv\sqrt{2}g_{0}\alpha_{j}=G_{j}^{x}+iG_{j}^{y}$, can be significantly enhanced by increasing the intracavity photons. Moreover, the solution of the linearized QLEs~(\ref{lqles}) is given by
\begin{align}
u(t)=M(t)u(0)+\int_{0}^{t}\textrm{d}\tau\, M(\tau)v(t-\tau),
\end{align}
where $M(t)=\exp(At)$. When all of the eigenvalues of the matrix $A$ have negative real parts, the system is stable and reaches its steady state, leading to $M(\infty)=0$ and
\begin{align}
u_{i}(\infty)=\int_{0}^{\infty}\textrm{d}\tau\,\sum_{k}M_{ik}(\tau)v_{k}(t-\tau).
\end{align}
The stability conditions can usually be derived by applying the Routh-Hurwitz criterion~\cite{dejesus1987RouthHurwitz}. In our following numerical simulations, we have confirmed that the chosen parameters in this paper can keep the COM system in a stable regime.

Because of the linearized dynamics of the QLEs and the Gaussian nature of the input noises, the steady state of the quantum fluctuations, independently of any initial conditions, can finally evolve into a tripartite continuous variable (CV) Gaussian state, which is fully characterized by a $6\times6$ stationary correlation matrix (CM) $V$ with the components

\begin{figure*}[t]
	\centering
	\includegraphics[width=\linewidth]{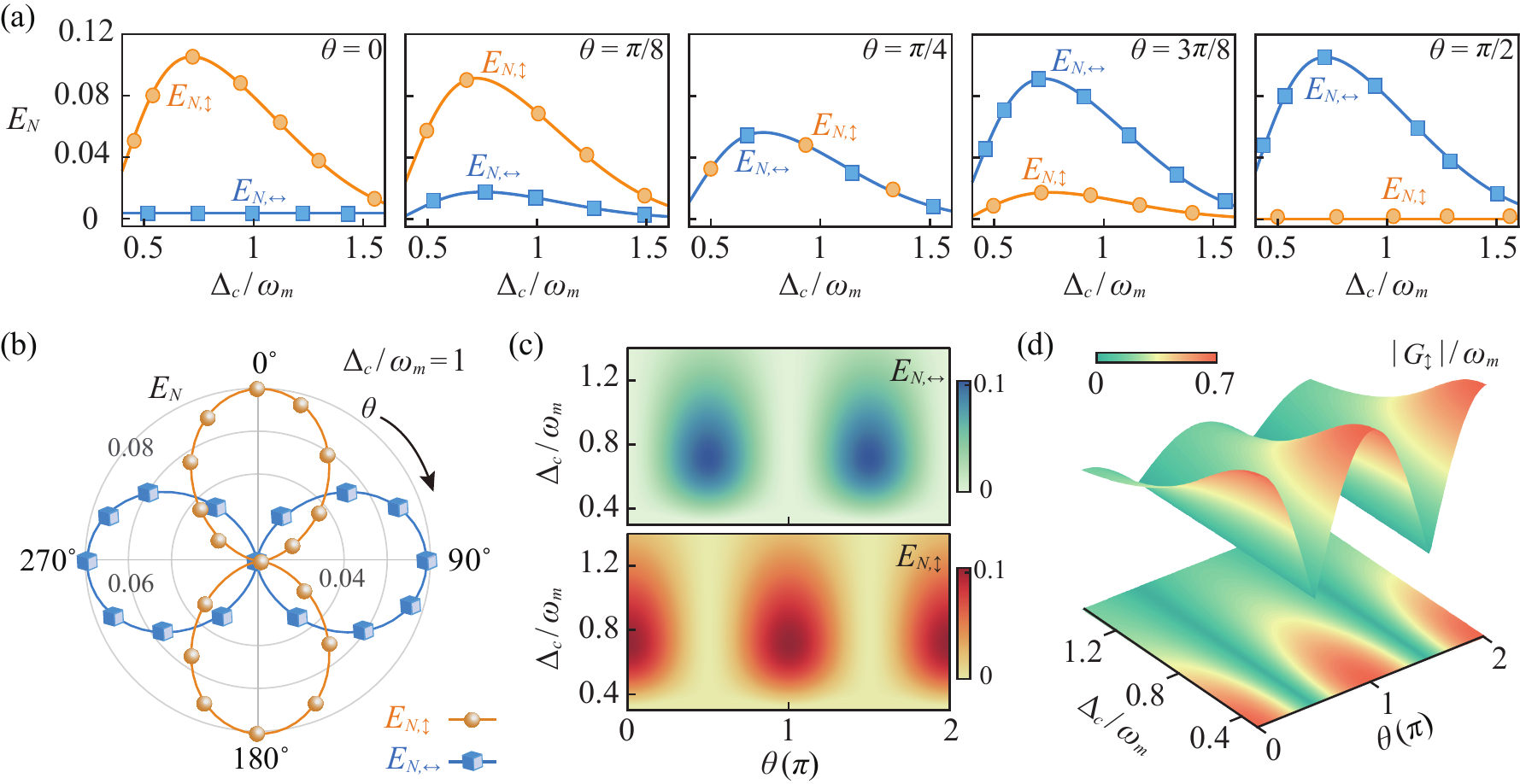}	
	\caption{Polarization-controlled optomechanical entanglement switch within intracavity mode.  (a) The logarithmic negativity $E_{N,j}$ of the TE and TM modes versus the scaled optical detuning $\Delta_{c}$ for different values of the polarization angle $\theta$. (b) The logarithmic negativity $E_{N,j}$ of the TE and TM modes versus the polarization angle $\theta$ in polar coordinates, with the optical detuning $\Delta_{c}/\omega_{m}=1$. $E_{N\!,\,\updownarrow}$ and $E_{N\!,\,\leftrightarrow}$ demonstrate a complementary distribution with the variation of the polarization angle $\theta$. (c) Density plot of $E_{N,j}$ as a function of the scaled optical detuning $\Delta_{c}$ and the polarization angle $\theta$. By tuning the polarization angle, an adjustable coherent COM entanglement switch between the TE and TM modes could be achieved. (d) The effective COM coupling $|G_{\updownarrow}|$ of the TE mode versus the scaled optical detuning $\Delta_{c}$ and the polarization angle $\theta$.}
	\label{FIG2}
\end{figure*}

\begin{align}
V_{ij}=\big\langle u_{i}(\infty)u_{j}(\infty)\!+\!u_{j}(\infty)u_{i}(\infty)\big\rangle/2.
\label{Vmatrix}
\end{align}
Using the solution of the steady-state QLEs, we can obtain the correlation matrix
\begin{align}
V=\int_{0}^{\infty}\textrm{d}\tau \,M(\tau)D M^{T}(\tau),
\label{eq26}
\end{align}
where
\begin{align}
D\!=\!\textrm{Diag}\,[\kappa,\kappa,\kappa,\kappa,0,\gamma_{m}(2n_{m}\!+\!1)],
\end{align}
is the diffusion matrix, which is defined through $\langle n_{i}(\tau)n_{j}(\tau')\!+\!n_{j}(\tau')n_{i}(\tau)\rangle/2\!=\!D_{ij}\delta(\tau-\tau')$.
When the stability condition is fulfilled, the dynamics of the steady-state correlation matrix is determined by the Lyapunov equation~\cite{vitali2007Optomechanical}:
\begin{align}
AV+VA^{\textit{T}}=-D.
\label{LyaV}
\end{align}
As seen from Eq.\,(\ref{LyaV}), the Lyapunov equation is linear and can straightforwardly be solved, thus allowing us to derive the correlation matrix $V$ for any values of the relevant parameters. However, the explicit form of $V$ is complicated and is not be reported here.

\section{Switching optomechanical entanglement by tuning the polarization of vector light}
To explore the polarization-controlled coherent switch of the steady-state COM entanglement, we adopt the logarithmic negativity, $E_{N}$, for quantifying the bipartite distillable entanglement between different degrees of freedom of our three-mode Gaussian state~\cite{adesso2004Extremal}. In the CV case, $E_{N}$ can be defined as
\begin{align}
E_{N}=\max\,\left[0,-\ln\left(2\tilde{\nu}_{-}\right)\right],
\label{En}
\end{align}
where
\begin{align}
\tilde{\nu}_{-}\!=\!2^{-1/2}\big\{\Sigma(V_{bp})-\big[\Sigma(V_{bp})^{2}-4\det\,V_{bp}\big]^{1/2}\big\}^{1/2},
\end{align}
with
\begin{align}
\Sigma(V_{bp})=\det\,A+\det\,B-2\det\,C.
\end{align}
Here $\tilde{\nu}_{-}$ is the minimum symplectic eigenvalue of the partial transpose of the reduced $4\times4$ CM $V_{bp}$. By tracing out the rows and columns of the uninteresting mode in $V$, the reduced CM $V_{bp}$ can be given in a $2\times2$ block form
\begin{align}
V_{bp}=\left(
\begin{matrix}
A&C\\
C^{\textit{T}}&B
\end{matrix}
\right).
\end{align}
Equation (\ref{En}) indicates that the COM entanglement emerges only when $\tilde{\nu}_{-}<1/2$, which is equivalent to the Simon's necessary and sufficient entanglement nonpositive partial transpose criterion (or the related Peres-Horodecki criterion) for certifying bipartite CV distillable entanglement in Gaussian states~\cite{simon2000PeresHorodecki}. Therefore, $E_{N}$, quantifying the amount by which the Peres-Horodecki criterion is violated, is an efficient entanglement measure that is widely used when studying bipartite entanglement in a multi-mode system.

In Figure\,\ref{FIG2}, the logarithmic negativity $E_{N\!,\,j}$ and the associated effective COM coupling rate $G_{j}$ with the intracavity field are shown as a function of the optical detuning $\Delta_{c}$ with respect to different polarization angle $\theta$. Here, $E_{N\!,\,\updownarrow}$ and $E_{N\!,\,\leftrightarrow}$ are used to denote the case of the COM entanglement with respect to the TE and TM modes, respectively. For ensuring the stability and experimental feasibility, the parameters are moderately chosen as follows: $m\!=\!50\,\textrm{ng}$, $\lambda\!=\!810\,\textrm{nm}$, $\omega_{m}/2\pi\!=\!10\,\textrm{MHz}$, $g_{0}/2\pi\!=\!68.5\,\textrm{Hz}$, $Q_{c}\!=\!\omega_{c}/\kappa\!=\!4.94\times10^{7}$, $Q_{m}\!=\!\omega_{m}/\gamma_{m}\!=\!10^{5}$, $\mathrm{T}\!=\!400\,\textrm{mK}$, and $P\!=\!30\,\textrm{mW}$. Note that $Q_{c}$ is typically $10^{5}-10^{10}$~\cite{aspelmeyer2014Cavity,verlot2009Scheme}, and $Q_{m}$ is typically $10^{5}-10^{6}$~\cite{galinskiy2020Phonon} in Fabry-P\'{e}rot cavities. Using a whispering-gallery-mode COM system, much higher value of $Q_{c}$ is achievable, reaching even up to $10^{12}$~\cite{huet2016Millisecond}. The polarization angle $\theta$, which can be tuned by rotating the orientations of the optical polarizer, is usually used to control the spatial amplitude and phase of a linearly polarized driving field, with values in a period ranging from $0$ to $2\pi$~\cite{mohdnasir2019Polarization}. In particular, $\theta=0$ ($\theta=\pi/2$) corresponds to the case with a vertically (horizontally) polarized pump field applied to drive the TE (TM) optical mode.	

\begin{figure}[t]
\centering
\includegraphics[width=0.42\textwidth,height=0.52\textheight]{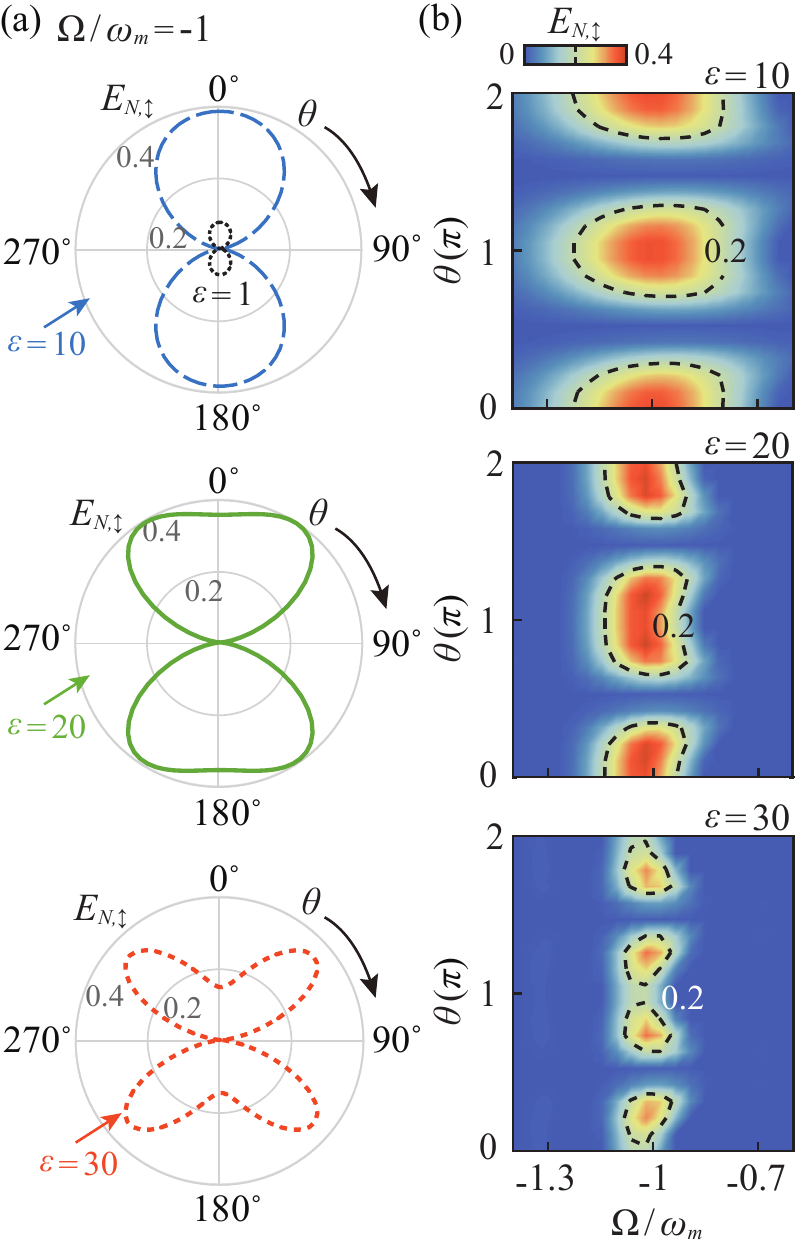}	
\caption{Polarization-controlled optomechanical entanglement within the cavity output mode. (a) The logarithmic negativity of the TE mode, $E_{N\!,\,\updownarrow}$ , as a function of the polarization angle $\theta$ in polar coordinates for different values of the inverse bandwidth $\varepsilon$, with the central frequency $\Omega/\omega_{m}=-1$. (b)  $E_{N\!,\,\updownarrow}$ versus the central frequency $\Omega$ and the polarization angle $\theta$ for different values of $\varepsilon$. There is an optimal value of $\theta$ for achieving the maximum COM entanglement within the cavity output mode.}
\label{FIG3}
\end{figure}

Specifically, as shown in Figure\,\ref{FIG2}(a)--(c), the COM entanglement is present only within a finite interval of the values of $\Delta_{c}$ around $\Delta_{c}\simeq\omega_{m}$, and the spectral offset of the logarithmic negativity peak is due to the radiation pressure induced redshift of the cavity mode (see, e.g., Ref.\,\cite{vitali2007Optomechanical, genes2008Robust}). In fact, the COM entanglement survives only with a tiny thermal noise occupancy. Therefore, at the light-motion resonance $\Delta/\omega_{m}\simeq1$, the COM interaction could significantly cool the mechanical mode and, simultaneously, leads to a considerable COM entanglement. In addition, the logarithmic negativities $E_{N\!,\,\leftrightarrow}$ and $E_{N\!,\,\updownarrow}$ always demonstrate a complementary distribution with the variation of the polarization angle $\theta$, indicating that an adjustable COM entanglement conversion between the TE and the TM modes would be implemented by a coherent polarization control. The underlying physics of this phenomenon can be understood as follows. In the polarization-controlled COM system, the field intensity of a the TE and TM modes is dependent on the spatial distribution of the linearly polarized driving field~\cite{wang2020Vectorial}. Correspondingly, the strength of the effective COM coupling rate $|G_{j}|$ now also relies on polarization angle $\theta$. As shown in Figure\,\ref{FIG2}(c) and \ref{FIG2}(d), it can be clearly seen that $E_{N\!,\,\updownarrow}$ achieves its maximum value for the optimal value of $|G_{\updownarrow}|$. Therefore, the distribution of the COM entanglement with respect to the TE and TM modes could be manipulated by tuning the polarization angle. In practical aspect, the ability to generate an adjustable entanglement conversion between subsystems of a compound COM system would provide another degree of freedom for quantum optomechanical information processing.

For practical applications, the COM entanglement with the intracavity field is hard to be directly accessed and utilized. In order to verify the generated COM entanglement, an essential step is to perform homodyne or heterodyne detections to the cavity output field, which allow to measure the corresponding CM $V^{\textrm{out}}$. Specifically, the quantum correlations in $V^{\textrm{out}}$, that involve optical quadratures, can be directly read out by homodyning the cavity output. However, accessing the mechanical quadratures typically requires to map the mechanical motion to a weak probe field first, which then can be read out by applying a similar homodyne procedure of the probe field.

Now, by applying the standard input-output relations, the output field of this compound COM system is given by
\begin{align}
a_{j}^{\mathrm{out}}(t)=\sqrt{2}\delta
a_{j}(t)-a_{j}^{\mathrm{in}}(t),
\label{inoutfun}
\end{align}
where the optical output field has the same non-zero correlation function as the input field $\delta a_{j}(t)$, i.e., $[a_{j}^{\mathrm{out}}(t)$, $a_{j}^{\mathrm{out}}(t')^{\dagger}]=\delta(t-t')$. As discussed in detail in Ref.\,\cite{vitali2008Timeseparated}, by selecting different time or frequency intervals from the continuous output field $a_{j}^{\mathrm{out}}(t)$, one can extract a set of independent optical modes by means of spectral filters. Here, for convenience, we consider the case where only a single output mode of the TE and TM cavity field is detected. Therefore, in terms of a causal filter function $g(s)$, the output field can be rewritten as
\begin{align}
a_{1j}^{\mathrm{out}}(t)=\int_{-\infty}^{t} ds \, g_{j}(t-s)a_{j}^{\mathrm{out}}(s).
\label{iorelation}
\end{align}
In the frequency domain, $a_{1j}^{\mathrm{out}}$ takes the following form
\begin{align}
\tilde{a}_{1j}^{\mathrm{out}}(\omega)=\int_{-\infty}^{\infty}\frac{dt}{\sqrt{2\pi}}a_{1j}^{\mathrm{out}}(t)\textrm{exp}(i\omega t)=\sqrt{2\pi}\tilde{g}_{j}(\omega)a_{1j}^{\mathrm{out}}(\omega),
\end{align}
where $\tilde{g}_{j}(\omega)$ is the Fourier transform of the filter function. An explicit example of an orthonormal set of the causal filter functions is given by~\cite{yan2019Entanglementa}
\begin{align}
g_{j}(t)=\frac{\Lambda(t)-\Lambda(t-\tau)}{\sqrt{\tau}}\textrm{exp}(-i\Omega t),
\label{gj}
\end{align}
and $g_{j}=g_{j}^{x}+ig_{j}^{y}$, where $g_{j}^{x}$ $(g_{j}^{y})$ is the real (imaginary) part of $g_{j}$, $\Lambda(t)$ denotes the Heaviside step function, with $\Omega$ and $\tau^{-1}$ being the central frequency and the bandwidth of the causal filter, respectively.
Here, $\tilde{g}(\omega)$ is given by
\begin{align}
\tilde{g}_{j}(\omega)=\sqrt{\frac{\tau}{2\pi}}\textrm{exp}[i(\omega-\Omega)\tau/2]\frac{\sin[(\omega-\Omega)\tau/2]}{(\omega-\Omega)\tau/2}.
\end{align}
The COM entanglement for the cavity output mode is verified via the CM defined as follows, i.e.,
\begin{align}
\mathrm{V}_{kl}^{\mathrm{out}}(t)=\frac{1}{2}\big\langle u_{k}^{\mathrm{out}}(t)u_{l}^{\mathrm{out}}(t)+u_{l}^{\mathrm{out}}(t)u_{k}^{\mathrm{out}}(t)\big\rangle,
\end{align}
where $u^{\mathrm{out}}(t)=(\delta X_{\updownarrow}^{\mathrm{out}}, \delta Y_{\updownarrow}^{\mathrm{out}}, \delta X_{\leftrightarrow}^{\mathrm{out}}, \delta Y_{\leftrightarrow}^{\mathrm{out}}, \delta q,\delta p)^{\textit{T}}$
is the vector form by the mechanical position and momentum fluctuations and by the (canonical) position quadrature,
\begin{align}
X_{j}^{\mathrm{out}}(t)=\big\{a_{1j}^{\mathrm{out}}(t)+\big[a_{1j}^{\mathrm{out}}(t)\big]^{\dagger}\big\}/\sqrt{2},
\end{align}
and the momentum quadrature,
\begin{align}
Y_{j}^{\mathrm{out}}(t)=\big\{a_{1j}^{\mathrm{out}}(t)-\big[a_{1j}^{\mathrm{out}}(t)\big]^{\dagger}\big\}/i\sqrt{2},
\end{align}
of the optical output modes. From the input-output relation in Equation (\ref{inoutfun}), one can obtain
\begin{align}
u_{i}^{\mathrm{out}}(t)=\int_{-\infty}^{t} ds\, \mathcal{T}_{i}(t-s)[u(s)- v'(s)],
\end{align}
where
\begin{align}
v'(t)&=\frac{1}{\sqrt{2\kappa}}(X_{\updownarrow}(t),Y_{\updownarrow}(t),X_{\leftrightarrow}(t),Y_{\leftrightarrow}(t),0,0)^{T},
\end{align}
is analogous to the noise vector $v(t)$ in Equation (\ref{lqles}) but without noise acting on the mechanical mode. We have also introduced the matrix,
\begin{align}
\mathcal{T}(t)=\left(\begin{smallmatrix}
f_{\updownarrow}^{x} & -f_{\updownarrow}^{y} & 0 & 0 & 0 & 0\\
f_{\updownarrow}^{y} & f_{\updownarrow}^{x}  & 0 & 0 & 0 & 0 \\
0 & 0 & f_{\leftrightarrow}^{x} & -f_{\leftrightarrow}^{y} & 0 & 0\\
0 & 0 & f_{\leftrightarrow}^{y} & f_{\leftrightarrow}^{x}  & 0  & 0\\
0 & 0 & 0 & 0 & \delta(t) & 0 &  \\
0 & 0 & 0 & 0 & 0 & \delta(t)
\end{smallmatrix}\right),
\end{align}
where $f_{j}^{x} = \sqrt{2\kappa}g_{j}^{x}, f_{j}^{y} = \sqrt{2\kappa}g_{j}^{y}$. Using the Fourier transform and the correlation function of the noises, one can derive the following general expression for the stationary output correlation matrix, which is the counterpart of the intracavity relation of Equation (\ref{eq26}):
\begin{align}
V^{\mathrm{out}}=&\int  d\omega\,\tilde{T}(\omega)\left[\tilde{M}(\omega)+\frac{P_{\mathrm{out}}}{2\kappa}\right] \notag \\
&\times D(\omega)\left[\tilde{M}^{\dagger}(\omega)+\frac{P_{\mathrm{out}}}{2\kappa}\right]\tilde{T}^{\dagger}(\omega),
\label{vout}
\end{align}
where $P_{\mathrm{out}}=\mathrm{Diag}[1,1,1,1,0,0]$ is the projector onto the two-dimensional space associated with the output quadratures, and
\begin{align}
\tilde{M}^{\mathrm{ext}}(\omega)=(i\omega+A)^{-1},
\end{align}
and
\begin{equation}
D(\omega)=\begin{pmatrix}
\kappa & 0 & 0 & 0 & 0 & 0 \\
0 & \kappa & 0 & 0 & 0 & 0\\
0 & 0 & \kappa & 0 & 0 & 0 \\
0 & 0 & 0 & \kappa & 0 & 0 \\
0 & 0 & 0 & 0 & 0 & 0\\
0 & 0 & 0 & 0 & 0 & N_{m}
\end{pmatrix},
\label{39}
\end{equation}
where $N_{m}=(\gamma_{m}\omega/\omega_{m})\coth(\hbar\omega/2k_{B}T)$. A deeper understanding of the general expression for $V^{\mathrm{out}}$ in Equation (\ref{vout}) is obtained by multiplying the terms in the integral, one obtains
\begin{align}
V^{\mathrm{out}}&=\int d\omega\,\tilde{T}(\omega)\tilde{M}^{\dagger}(\omega)D(\omega)\tilde{M}(\omega)\tilde{T}^{\dagger}(\omega)+\frac{P_{\mathrm{out}}}{2}\nonumber\\&\quad+\int d\omega\,\tilde{T}(\omega)\big[\tilde{M}(\omega)R_{\mathrm{out}}+R_{\mathrm{out}}\tilde{M}^{\dagger}(\omega)\big]\tilde{T}^{\dagger}(\omega),
\label{Vout}
\end{align}
where
$R_{\mathrm{out}}=P_{\mathrm{out}}D(\omega)/\kappa=D(\omega)P_{\mathrm{out}}/\kappa$, for simplicity, we have defined~\cite{genes2008Robust}
\begin{align}
P_{\mathrm{out}}\equiv 2\int\frac{d\omega}{4\kappa^{2}}\tilde{T}(\omega)P_{\mathrm{out}}D(\omega)P_{\mathrm{out}}\tilde{T}^{\dagger}(\omega).
\end{align}
Equation (\ref{Vout}) clearly shows the quantum correlations of the quadrature operators in the output field, where the first integral term stems from the intracavity COM interaction, the second term gives the contribution the correlations of the noise operators, and the last term describes the interactions between the intracavity mode and the optical input field.

\begin{figure}[t]
	\centering	
    \includegraphics[width=0.43\textwidth]{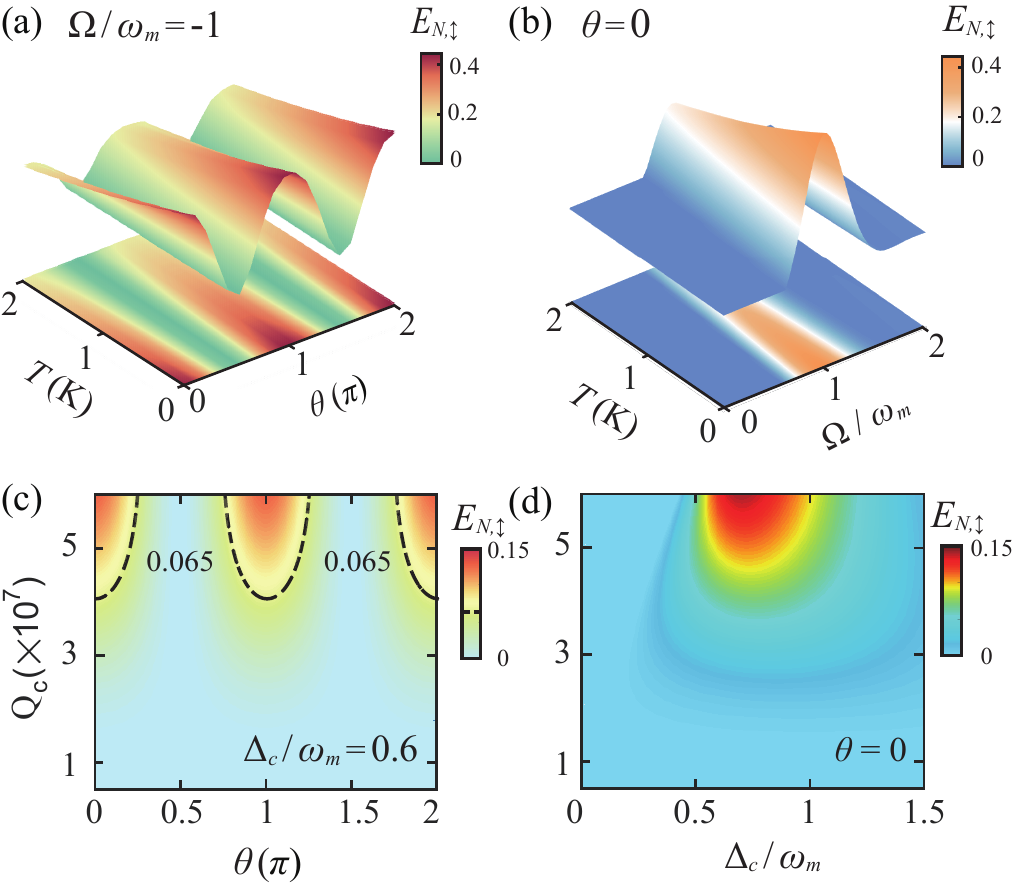}	
	\caption{ The role of thermal effects and quality factors on the generation and manipulation of COM entanglement. (a) The logarithmic negativity of the TE mode, $E_{N\!,\,\updownarrow}$ , as a function of the environment temperature $T$ and the polarization angle $\theta$, with the central frequency $\Omega/\omega_{m}=-1$ and $\varepsilon=10$. (b) $E_{N\!,\,\updownarrow}$  versus the central frequency $\Omega$ and the environment temperature $T$, with the polarization angle $\theta=0$ and $\varepsilon=10$. (c) Density plot of $E_{N\!,\,\updownarrow}$ versus the polarization angle $\theta$ and the quality factor $Q$, with the optical detuning $\Delta_{c}/\omega_{m}=0.6$. (d) $E_{N\!,\,\updownarrow}$ as a function of the scaled optical detuning $\Delta_{c}$ and the quality factor $Q$, with $\theta=0$}.
	\label{FIG4}
\end{figure}

Then, by numerically calculating the CM $V^{\mathrm{out}}$ and the associated logarithmic negativity $E_{N\!,\,j}$, one can detect and verify the generated intracavity COM entanglement at the cavity output. Here, as a specific example shown in Figure\,\ref{FIG3}, we studied the COM entanglement between the TE output mode and mechanical mirror. Also, as discussed in detail in Ref.\,\cite{genes2008Robust}, the generated intracavity COM entanglement is mostly carried by the lower-frequency Stokes sideband of the output field. Therefore, it would be better to choose an input field with the center frequency at $\Omega=-\omega_{m}$, which can usually be implemented by using the filter function of Equation (\ref{gj}). Specifically, for $\varepsilon=1$, as shown in Figure\,\ref{FIG3}(a), the variation of COM entanglement with the polarization angle at the cavity output field is similar to that of its intracavity counterpart. However, by optimizing the value of $\varepsilon$ or, equivalently, the detection bandwidth $\tau$, $E_{N\!,\,\updownarrow}$ would achieve much higher value than that of the intracavity field, implying a significant enhancement of COM entanglement at the output field. In addition, for particular detection bandwidth, there is also an optimal value of the polarization angle for which the COM entanglement of the output field achieves its maximum value [see Figure\,\ref{FIG3}(b) for more details]. The physical origin of the enhancement of COM entanglement at cavity output field results from the formation of quantum correlations between the intracavity mode and the optical input field may cancel the destructive effects of the input noises. The process of detecting entanglement at the output field is related to entanglement distillation, whose basic idea is originating from extracting, from an ensemble of pairs of non-maximally entangled qubits, a smaller number of pairs with a higher degree of entanglement~\cite{kwiat2001Experimental}. In this situation, it is possible that one can further manipulate and optimize the polarization-controlled COM entanglement at the output field, which can be useful for quantum communications~\cite{briegel1998Quantum,Xu2020}, quantum teleportation~\cite{Llewellyn2019}, and loophole-free Bell test experiment~\cite{garcia-patron2004Proposal}.

Besides, as discussed in details in Ref.\,\cite{genes2008Robust}, the thermal phonon excitations could greatly affect the quantum correlations between photons and phonons, and, thus, tend to destroy the generated COM entanglement. The mean thermal phonon number, which is monotonically increasing with the environment temperature, is assumed to be $n_{m}\simeq833$ in our former discussions. Here, by computing $E_{N\!,\,\updownarrow}$ as a function of temperature $T$, as shown in Figure\,\ref{FIG4}, we confirmed that the conversion of COM entanglement through polarization control could still exist when considering larger mean thermal phonon numbers. To clearly see this phenomenon, we plot the logarithmic $E_{N\!,\,\updownarrow}$ of the CV bipartite system formed by the mechanical mode and the cavity output mode centered around the Stokes sideband $\Omega=-\omega_{m}$ versus polarization angle $\theta$ and temperature $T$ in Figure\,\ref{FIG4}(a) and set $\theta=0$. We also consider $E_{N\!,\,\updownarrow}$ as a function of temperature $T$ and the cavity frequency $\Omega$ in Figure\,\ref{FIG4}(b). It is seen that for $\varepsilon=10$, the polarization-controlled COM entanglement can still be observed at the cavity output field below a critical temperature $T_{c}\approx2\,\textrm{K}$, corresponding to $n_{m}\simeq4160$. Furthermore, the quality factor $Q$ of the cavity mode could also affect the implementation of the COM entanglement switch. As shown in Figure 4(c)-4(d), for a certain environment temperature, e.g., $T=400\,\textrm{mK}$, the degree of COM entanglement is suppressed by decreasing the value of $Q$, and it is seen that the minimum $Q$ factor required for the COM entanglement switch process is approximately $Q\simeq3\times10^{7}$.

Finally, we remark that based on experimentally feasible parameters, not only all bipartite entanglements but also the genuine tripartite entanglement can be created and controlled in our system. In fact, we have already confirmed that the tripartite entanglement, measured by the minimum residual contangle~\cite{Adesso2007,Li2018}, can reach its maximum at the optimal polarization angle $\theta=\pi/4$. However, the bipartite purely optical entanglement of TE and TM modes is typically very weak due to the indirect coupling of these two optical modes, and thus the resulting tripartite entanglement is also weak in this system. Nevertheless, we note that the techniques of achieving stationary entanglement between two optical fields and even strong tripartite entanglement are already well available in current COM experiments~\cite{Chen2020,barzanjeh2019Stationary,Lepinay2021,OckeloenKorppi2018,Riedinger2018}, and our work here provides a complementary way to achieve coherent switch of the COM entanglement via polarization control.

\section{Conclusion}
In summary, we have proposed how to manipulate the light-motion interaction in a COM resonator via polarization control, which enables the ability of coherently switching COM entanglement in such a device. We note that the ability to achieve coherent switch of COM entanglement is useful for a wide range of entanglement-based quantum technologies, such as quantum information processing~\cite{PhysRevLett.109.013603}, quantum routing~\cite{Yuan2020}, or quantum networking~\cite{Kimble2008}. Also, our work reveals the potential of engineering various quantum effects by tuning the optical polarizations, such as mechanical squeezing~\cite{Wollman2015}, quantum state transfer~\cite{Kurpiers2018}, and asymmetric Einstein-Podolsky-Rosen steering~\cite{Armstrong2015}. Although we have considered here a specific case of linearly polarized field whose spatial distribution is homogeneous, we can envision that future developments with inhomogeneous vector beams can further facilitate more appealing quantum COM techniques, such as opto-rotational entanglement~\cite{jing2011Quantum} or polarization-tuned topological energy transfer~\cite{xu2019Nonreciprocal}. In a broader view, our findings shed new lights on the marriage of vectorial control and quantum engineering, opening up the way to control quantum COM states by utilizing synthetic optical materials~\cite{aspelmeyer2014Cavity,Chen2021}.

\textbf{Author contribution:}
 All the authors have accepted responsibility for the entire content of this submitted manuscript and approved submission.

\textbf{Research  funding:}
This work is supported by the National Natural Science Foundation of China (Grants No. 11935006 and No. 11774086) and the Science and Technology Innovation Program of Hunan Province (Grant No. 2020RC4047). L.-M. K. is supported by the National Natural Science Foundation of China (NSFC, Grants No. 11775075 and No. 11434011). A. M. was supported by the Polish National Science Centre
(NCN) under the Maestro Grant No. DEC-2019/34/A/ST2/00081.

\textbf{Conflict of interest statement:}
The authors declare no conflicts of interest regarding this article.


\end{document}